\documentclass{aip-cp}

\usepackage[numbers]{natbib}
\usepackage{rotating}
\usepackage{graphicx}
\usepackage{siunitx}
\usepackage{color}

\begin{document}

\title{The TeV supernova remnant shell HESS J1731-347 and its surroundings}

\author[aff1]{M. Capasso\corref{cor1}}
\author[aff2]{B. Condon}
\author[aff1]{M. Coffaro}
\author[aff1]{Y. Cui}
\author[aff1]{D. Gottschall}
\author[aff1]{D. Klochkov}
\author[aff3]{V. Marandon}
\author[aff4]{N. Maxted}
\author[aff1]{G. P\"uhlhofer}
\author[aff5]{G. Rowell}
\author[]{on behalf of the H.E.S.S. Collaboration}

\affil[aff1]{Institute for Astronomy and Astrophysics T\"ubingen}
\affil[aff2]{Universit\'e de Bordeaux, CNRS IN2P3, CENBG}
\affil[aff3]{Max-Planck-Institut f\"ur Kernphysik, Heidelberg}
\affil[aff4]{University of New South Wales, Australia}
\affil[aff5]{University of Adelaide, Australia}

\corresp[cor1]{Corresponding author: capasso@astro.uni-tuebingen.de}

\maketitle

\begin{abstract}
HESS\,J1731-347 is a shell-type supernova remnant emitting both TeV gamma rays and non-thermal X-ray photons, spatially coincident with the radio SNR G353.6-0.7. Hadronic and leptonic scenarios (or a blend of both) are discussed in the literature to explain the TeV emission from the object. In 2011, a $\gamma$-ray excess was also found in the neighborhood of the source (HESS J1729-345). Here we present results of an updated analysis obtained with the meanwhile available additional H.E.S.S. data. Beyond HESS\,J1731-347, the analysis reveals the morphology of the emission of the adjacent TeV source HESS J1729-345 and the emission in between the two sources in greater detail. The results permit us to correlate the TeV emission outside of the SNR with molecular gas tracers, and to confront the data with scenarios in which the TeV emission outside the SNR is produced by escaping cosmic rays.
\end{abstract}

\section{Introduction}
Supernova remnants (SNRs) are considered prime candidates for the efficient acceleration of Galactic Cosmic Rays (CRs) at their shock fronts, up to the so-called \textit{knee} in the CR spectrum ($\sim10^{15}\mathrm{eV}$) \citep{2008ApJ...678..939Z}, though direct observational evidence is still lacking.

For young TeV-bright SNRs (e.g.: RXJ\,1713.7-3946 and Vela Jr.) it might occur that the more energetic CRs are able to directly escape from the acceleration region \citep{2007ApJ...665L.131G}. 

One signature of this process would be the different morphology of the X-ray emission (dominated by electron synchrotron radiation) and the very high energy $\gamma$-ray emission; namely the latter would have a larger extension. This emission could also be enhanced in the presence of target material for $\pi^{0}$-production, such as dense gas clouds. As shown in the following, the newly available Mopra CS data seem to show a good correspondence between the gas density for a specific distance solution and the very high energy $\gamma$-ray emission, suggesting that the latter could be caused by CRs escaping the HESS\,J1731-347 shell and interacting with a nearby molecular cloud.
\section{Data analysis}
\begin{figure}[h!]
  \centerline{\includegraphics[width=0.5\textwidth]{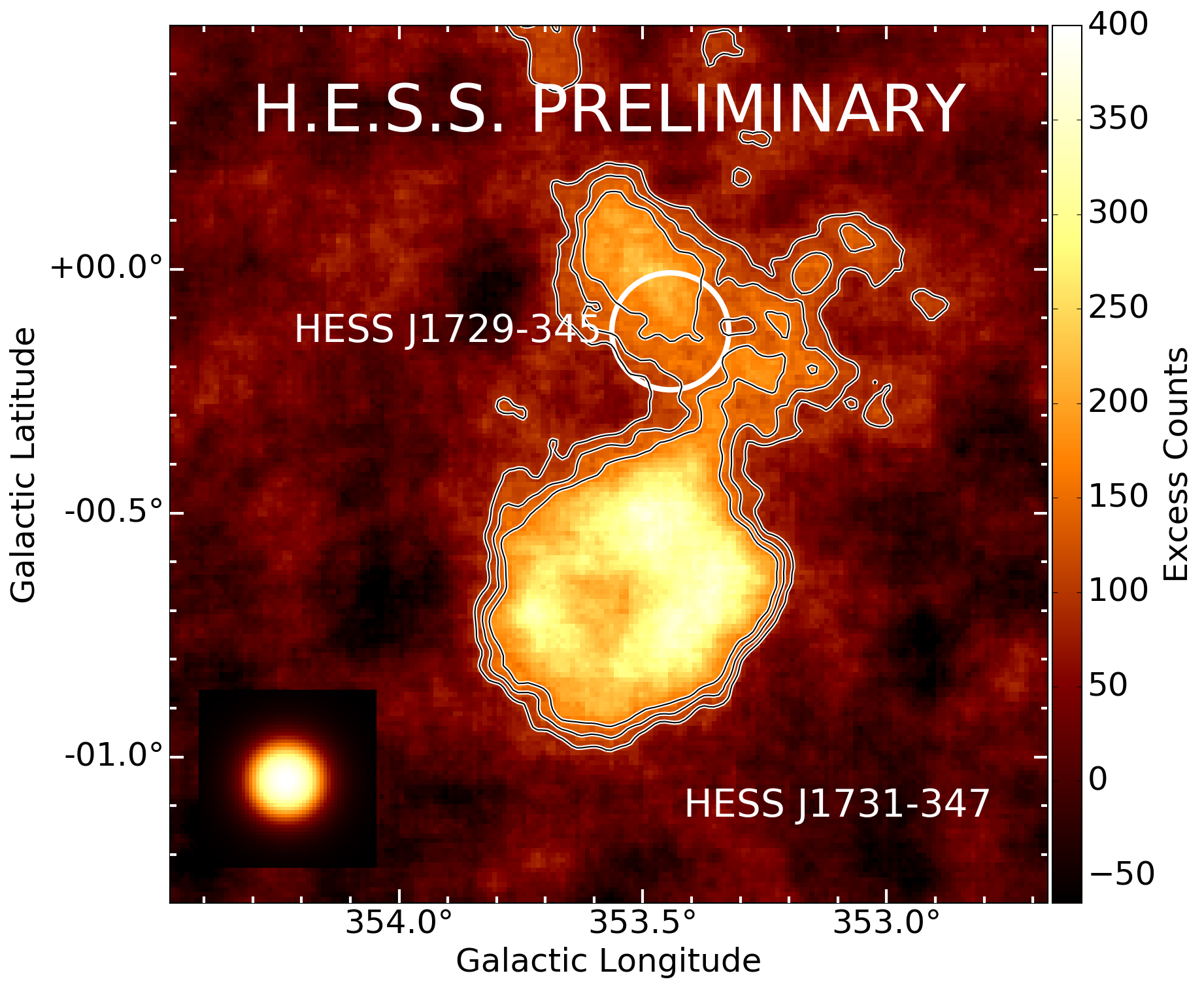}}
  \caption{$\gamma$-ray excess counts map of the HESS\,J1731-347 region, correlated with a circular filter with 0.1° radius. Overlaid: 3,4,5 sigma significance contours; the white circle shows the position and extension of HESS\,J1729-347 as reported in \citep{2011A&A...531A..81H}. The inlet on the bottom left shows the average PSF for this data set.}
  \label{fig:J1731}
\end{figure}

The data presented in this work have been taken with the H.E.S.S. (High Energy Sterescopic System) array\footnote{H.E.S.S. is a system of five Imaging Atmospheric Cherenkov Telescopes, located in the Khomas Highland of Namibia. H.E.S.S. phase I started in 2003 with four 12-m telescopes. A much larger 28-m fifth telescope has been operational since July 2012. In the presented work no data recorded with the fifth telescope are used.}, in the period between May 2004 and June 2013. Event direction and energy reconstruction are performed using a moment-based Hillas analysis as described in \citep{Aharonian:2006}, applying a cut of 60 photoelectrons (p.e.) on the image amplitude (\textit{standard} cuts). $\gamma$-ray like events are selected based on the image shapes with a boosted decision tree method \citep{Ohm:2009}. In order to produce the sky map of the source region (shown in Figure \ref{fig:J1731}), the residual background at each pixel is estimated from a ring around the source position using an adaptive algorithm to optimise the size of the ring, blanking out known sources or excesses above a certain significance level from the rings \citep{2007A&A...466.1219B}. The total observation time for HESS\,J1731-347 is $\sim75$h.

\section{Source distance estimates and escaping cosmic rays}
\begin{figure}[!htb]
\label{fig:gas}
\minipage{0.32\textwidth}
  \includegraphics[width=\linewidth,height=6cm]{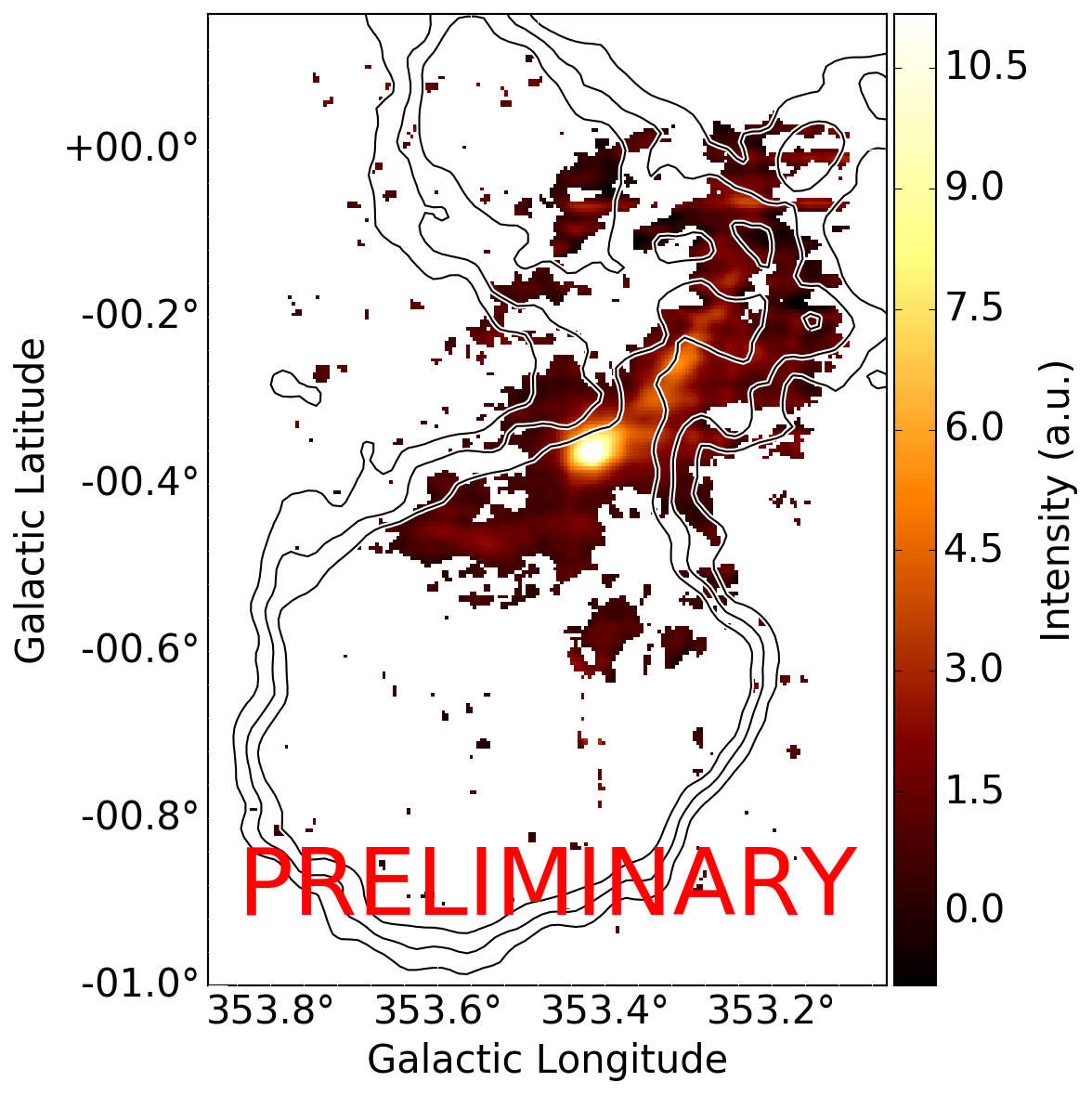}
\endminipage\hfill
\minipage{0.32\textwidth}
  \includegraphics[width=\linewidth,height=6cm]{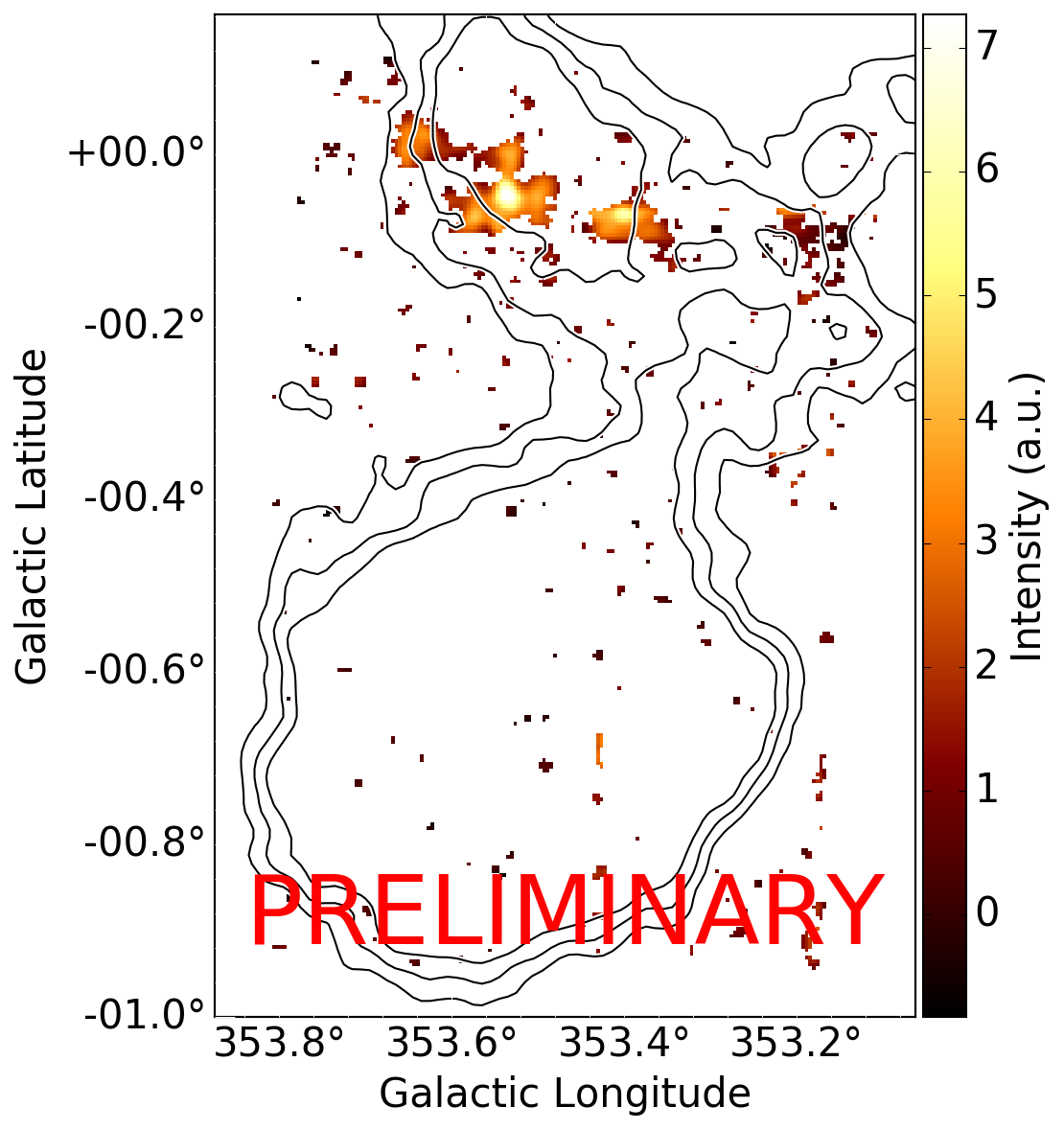}
\endminipage\hfill
\minipage{0.32\textwidth}
  \includegraphics[width=\linewidth,height=6cm]{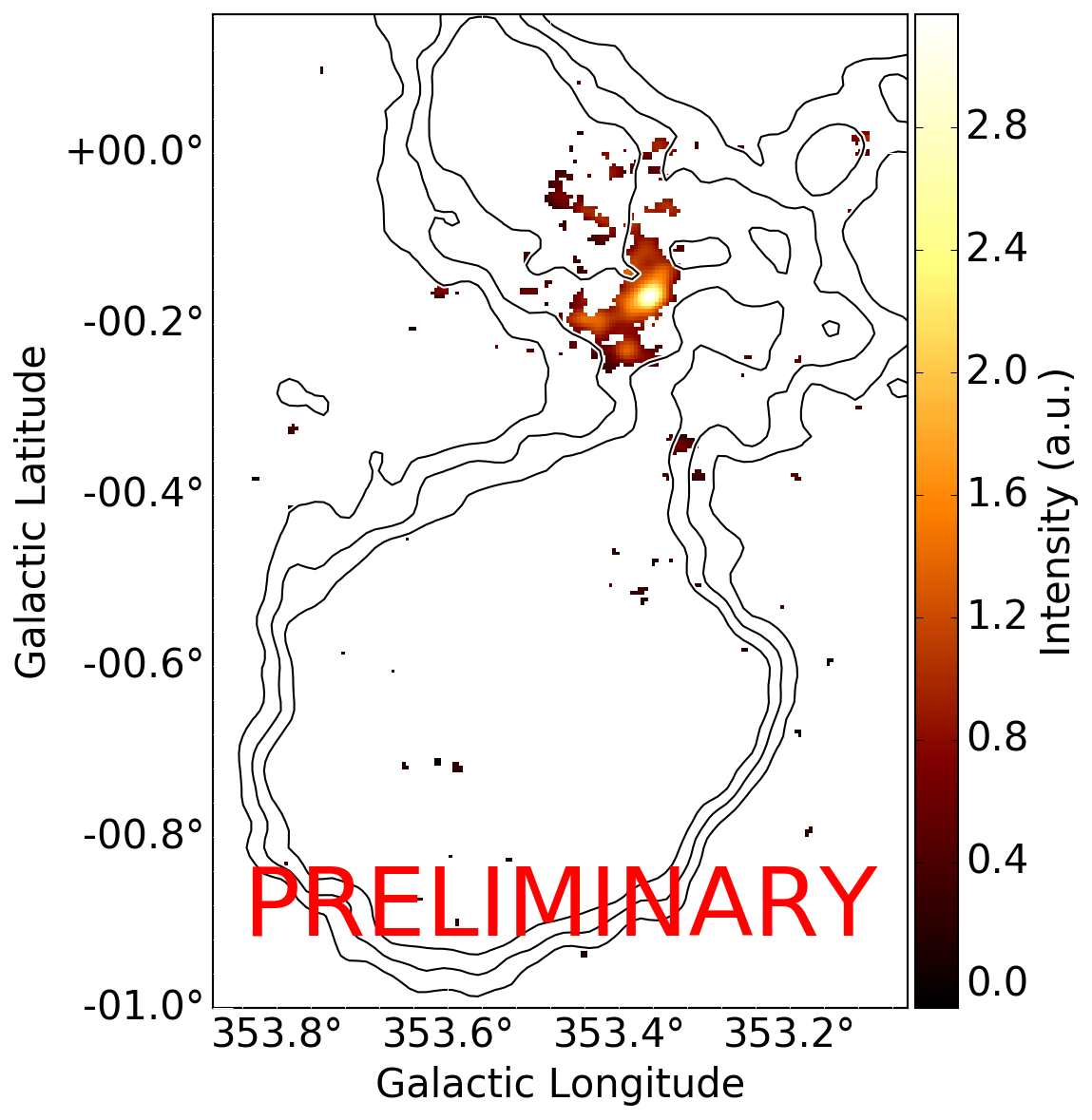}
\endminipage
\caption{The figure shows the integrated intensity maps in the HESS\,J1731-347 region for the three different velocity ranges in which a signal from CS is detected \citep{2015arXiv150306717M}. \textit{Left}: -23.7 to -7 km/s ($\sim3.2$ kpc) \textit{Middle}: -65 to -35 km/s ($\sim4.5$ kpc) \textit{Right}: -85.5 to -78 km/s ($\sim5-6$ kpc). Overlaid 3,4,5 significance contours from the H.E.S.S. analysis of the source presented in this work.}
\end{figure}

By comparing the interstellar absorption derived from X-rays and the one obtained from $^{12}\mathrm{CO}$ and $\mathrm{HI}$ observations, a lower limit on the source distance of $\sim3.2$ \SI{}{kpc} (Scutum-Crux arm) is proposed in \citep{2011A&A...531A..81H}. However, a farther distance of $\sim4.5$ \SI{}{kpc} (Norma arm) is not excluded.
Fukuda et al. \citep{2014ApJ...788...94F} analyse the interstellar protons towards the SNR by using both the $^{12}\mathrm{CO}$ and $\mathrm{HI}$ data sets, finding a spatial correlation in the velocity range from $-90$ \SI{}{km/s} to $-75$ \SI{}{km/s} (which corresponds to the 3 kpc expanding arm and places the SNR at a distance between 5.3 and 6.1 kpc). Klochkov et al. \citep{2015A&A...573A..53K} use numerical spectral models for carbon and hydrogen atmospheres to fit the spectrum of the central compact object (CCO) located towards the geometrical center of the SNR, supporting the possibility of the source located in the Scutum-Crux arm. In a recent paper, Cui et al. \citep{2016A&A...591A..68C} demonstrate that, assuming HESS J1731-347 is located at a distance of 3.2 kpc, it is possible that the emission seen in TeV from HESS\,J1729-345 is produced by cosmic rays escaping from the SNR and illuminating nearby molecular coulds. \\
Recently, the Mopra radio telescope was used to carry out a survey of the Galactic Plane in the CO(1-0) transition. Along with CO, CS data have been taken, tracing densities of $\sim10^{4} \mathrm{cm^{-3}}$ \citep{2015arXiv150306717M} (see Figure \ref{fig:gas}).

\section{Conclusions}
As can be seen from the maps shown in Figure \ref{fig:gas}, there seems to be a good spatial correspondence between the CS(1-0) intensity, an indicator of dense gas column density, seen at 3.2 kpc and the TeV emission in the \textit{bridge} region connecting HESS\,J1731-347 and HESS\,J1729-345. On the other hand, the limited correspondence in the other two ranges does not seem to support the association between gas and TeV emission for farther distances; studies to try and quantify numerically the correlation are ongoing.\\
If confirmed, the association between TeV and gas emission in the first velocity range would support the argument that the SNR is indeed located at a distance of 3.2 kpc and that the TeV emission observed in the surroundings of the source could be explained by cosmic ray particles that escape from the SNR and illuminate a nearby molecular cloud.

\section{ACKNOWLEDGMENTS}
The support of the Namibian authorities and of the University of Namibia in facilitating the construction and operation of H.E.S.S. is gratefully acknowledged, as is the support by the German Ministry for Education and Research (BMBF), the Max Planck Society, the German Research Foundation (DFG), the French Ministry for Research, the CNRS-IN2P3 and the Astroparticle Interdisciplinary Programme of the CNRS, the U.K. Science and Technology Facilities Council (STFC), the IPNP of the Charles University, the Czech Science Foundation, the Polish Ministry of Science and Higher Education, the South African Department of Science and Technology and National Research Foundation, the University of Namibia, the Innsbruck University, the Austrian Science Fund (FWF), and the Austrian Federal Ministry for Science, Research and Economy, and by the University of Adelaide and the Australian Research Council. We appreciate the excellent work of the technical support staff in Berlin, Durham, Hamburg, Heidelberg, Palaiseau, Paris, Saclay, and in Namibia in the construction and operation of the equipment. This work benefited from services provided by the H.E.S.S. Virtual Organisation, supported by the national resource providers of the EGI Federation.


\nocite{*}
\bibliographystyle{aipnum-cp}%
\bibliography{bibliography}%

\end{document}